# Faddeev-Jackiw Quantization of the Gauge Invariant Self-dual Fields Relative to String Theory

Leng Liao [1]     Yong-Chang Huang [1,2; 1)]

1 (Institute of Theoretical Physics, Beijing University of Technology, Beijing 100022, China )

2 (CCAST (World Lab.), P. O. Box 8735, Beijing 100080, China )

**Abstract:** We obtain a new symplectic Lagrangian density and deduce Faddeev-Jackiw (FJ) generalized brackets of the gauge invariant self-dual fields interacting with gauge fields. We further give FJ quantization of this system. Furthermore, the FJ method is compared with Dirac method, the results show the two methods are equivalent in the quantization of this system. And by the practical research in this letter, it can be found that the FJ method is really simpler than the Dirac method, namely, the FJ method obviates the need to distinguish primary and secondary constraints and first- and second-class constraints. Therefore, the FJ method is a more economical and effective method of quantization.

# 1 Introduction

The systems described by singular Lagrangians are called singular systems, and this kind of systems contains inherent constraints[1,2]. Electromagnetic field theory[1,2] and Yang-Mills field theory[1,2] are both singular systems, and in many fields of physics, there exist singular systems, such as gravitational field theory, supersymmetry theory, supergravitation theory, superstring theory etc. The investigation on inherent constraints has become as one basic task of the theoretical research on these theories.

---

1)   Corresponding author

The study of singular systems was started by Dirac[3]. Dirac proposed a kind of bracket (now called as Dirac bracket) to quantize singular systems. The quantization method used in this paper is Faddeev-Jackiw (FJ) quantization method[4–8]. In contrast with Dirac method, FJ method has the advantages of simplicity, obviates the need to distinguish primary and secondary constraints and first- and second-class constraints. Moreover, in FJ method, there is no the hypothesis of Dirac's conjecture. So, after it has been proposed, many physicists paid close attention to FJ method.

Ref.[4] researched the self-dual theory in terms of FJ method, the quantization of self-dual fields may be relative to string theory[9], Ref.[10] found that N = 2 (4) string theory is self-dual N = 4 Yang-Mills theory, and Ref.[11] researched that Non-BPS brane dynamics and dual tensor gauge theory, up to now, it evokes much attention[12]. There have been many techniques to construct the interaction theories between self-dual fields and gauge fields[13-16], but those theories all have some flaws. In this paper, we choose the coupling Lagrangian density of Ref.[17], which not only is a Lorentz invariant theory, but also is gauge invariant. Thus, it is obviously much better than the previous theories.

## 2 Faddeev-Jackiw quantization of gauge invariant self-dual fields

Gauge invariant self-dual fields interacting with gauge fields are described by the Lagrangian density[17]

$$\mathcal{L} = \dot{\phi}\phi' - (\phi')^2 + \frac{1}{2}(\dot{A}_1 - A_0')^2 + c\phi'(A_0 - A_1) - \frac{1}{2}c^2 A_1^2 - \dot{\theta}\theta' - (\theta')^2 + c\theta'(A_0 + A_1), \quad (1)$$

where Eq. (1) is a Lagrangian density in the (1+1) spacetime. It can be found that Eq. (1) is not a first-order Lagrangian density. So, before making the FJ process, it must be transformed into first-order Lagrangian density by introducing auxiliary fields. Here, the canonical momenta are chosen to be auxiliary fields.

The canonical momenta are given as follows

$$\pi^0 = \frac{\partial}{\partial \dot{A}_0}\mathcal{L} = 0, \ \pi^1 = \frac{\partial}{\partial \dot{A}_1}\mathcal{L} = \dot{A}_1 - A_0', \ \pi_\phi = \frac{\partial}{\partial \dot{\phi}}\mathcal{L} = \phi', \ \pi_\theta = \frac{\partial}{\partial \dot{\theta}}\mathcal{L} = -\theta'. \quad (2)$$

Correspondingly, the canonical Hamiltonian density is

$$\mathcal{H}c = \frac{1}{2}(\pi^1)^2 + \pi^1 A_0' + (\phi')^2 - c\phi'(A_0 - A_1) + \frac{1}{2}c^2 A_1^2 + (\theta')^2 - c\theta'(A_0 + A_1).$$

The first-order symplectic Lagrangian density is given by

$$\mathcal{L} = \xi_5 \dot{\xi}_2 + \xi_3' \dot{\xi}_3 - \xi_4' \dot{\xi}_4 - V(\xi), \quad (3)$$

where $\xi_1 = A_0$, $\xi_2 = A_1$, $\xi_3 = \phi$, $\xi_4 = \theta$, $\xi_5 = \pi^1$, which are symplectic coordinates, and $V(\xi) = \mathcal{H}c(\xi)$, namely

$$\mathcal{H}c(\xi) = \frac{1}{2}(\xi_5)^2 + \xi_5 \xi_1' + (\xi_3')^2 - c\xi_3'(\xi_1 - \xi_2) + \frac{1}{2}c^2 \xi_2^2 + (\xi_4')^2 - c\xi_4'(\xi_1 + \xi_2). \quad (4)$$

The components of symplectic 1-forms are

$a_1 = 0$, $a_2 = \xi_5$, $a_3 = \xi_3'$, $a_4 = -\xi_4'$, $a_5 = 0$. Using $f_{ij} = \frac{\delta a_j(y)}{\delta \xi_i(x)} - \frac{\delta a_i(x)}{\delta \xi_j(y)}$ $(i, j = 1,...,5)$,

we obtain the symplectic matrix

$$(f_{ij}) = \begin{pmatrix} 0 & 0 & 0 & 0 & 0 \\ 0 & 0 & 0 & 0 & -1 \\ 0 & 0 & -2\partial_x & 0 & 0 \\ 0 & 0 & 0 & 2\partial_x & 0 \\ 0 & 1 & 0 & 0 & 0 \end{pmatrix} \delta(x-y), \quad (5)$$

which is obviously singular. The zero-mode of this matrix is $(v^{(0)})^T = (v(x) \ 0 \ 0 \ 0 \ 0)$, where $\upsilon(x)$ is an arbitrary function. According to FJ method[6], using the zero-mode we can get the primary constraint is obtained as

$$\Omega^{(0)} = \xi_5'(x) + c[\xi_3'(x) + \xi_4'(x)] = 0 \quad . \tag{6}$$

And the first-iterated Lagrangian density is given as follows[6]

$$\mathcal{L}^{(1)} = \xi_5 \dot{\xi}_2 + \xi_3' \dot{\xi}_3 - \xi_4' \dot{\xi}_4 + \Omega^{(0)} \dot{\lambda} - V^{(1)}(\xi), \tag{7}$$

where $\dot{\lambda}$ is a function of time and space, and is a derivative of time, which does not depend on $\xi$, and then we have

$$V^{(1)} = V|_{\Omega^{(0)}=0} = \frac{1}{2}(\xi_5)^2 + \xi_5 \xi_1' + (\frac{1}{c}\xi_5' + \xi_4')^2 + (\xi_5' + c\xi_4')(\xi_1 - \xi_2) \\ + \frac{1}{2}c^2\xi_2^2 + (\xi_4')^2 - c\xi_4'(\xi_1 + \xi_2) \tag{8}$$

Continuing FJ process, and defining the first-iterated symplectic coordinates $\xi^{(1)} = (\xi, \lambda)$, we obtain the components of first-iterated symplectic 1-forms $a^{(1)}_1 = 0$, $a^{(1)}_2 = \xi_5$, $a^{(1)}_3 = \xi_3'$, $a^{(1)}_4 = -\xi_4'$, $a^{(1)}_5 = 0$, $a^{(1)}_6 = \Omega^{(0)}$, so the first-iterated symplectic matrix is deduced

$$(f_{ij}^{(1)}) = \begin{pmatrix} 0 & 0 & 0 & 0 & 0 & 0 \\ 0 & 0 & 0 & 0 & -1 & 0 \\ 0 & 0 & -2\partial_x & 0 & 0 & -c\partial_x \\ 0 & 0 & 0 & 2\partial_x & 0 & -c\partial_x \\ 0 & 1 & 0 & 0 & 0 & -\partial_x \\ 0 & 0 & -c\partial_x & -c\partial_x & -\partial_x & 0 \end{pmatrix} \delta(x-y). \tag{9}$$

where c is a constant. Obviously, this matrix is still a singular matrix, which has two zero-modes. By using the two zero modes, no new constrains can be got, however, the symplectic matrix is singular yet. So this system has gauge symmetries, we must introduce gauge conditions. Here, we choose gauge conditions $\Omega_1 = -\xi_1 - \xi_2 = 0$, $\Omega_2 = \xi_4' = 0$ [17] to fix gauges. Considering the two gauge conditions as constrains, according FJ method[6], a new symplectic Lagrangian density is constructed as

$$\mathcal{L}^{(2)} = \xi_5 \dot{\xi}_2 + \xi_3' \dot{\xi}_3 - \xi_4' \dot{\xi}_4 + \Omega^{(0)} \dot{\lambda} + \Omega_1 \dot{\eta}_1 + \Omega_2 \dot{\eta}_2 - V^{(2)}(\xi), \tag{10}$$

similarly, where $\eta_1$ and $\eta_2$ are functions of time and space, and which do not depend on $\xi^{(1)}$. The new symplectic coordinates and components of symplectic 1-forms are given as $\xi^{(2)} = (\xi, \lambda, \eta_1, \eta_2)$ and $a^{(2)}{}_1 = 0$, $a^{(2)}{}_2 = \xi_5$, $a^{(2)}{}_3 = \xi'_3$, $a^{(2)}{}_4 = -\xi'_4$, $a^{(2)}{}_5 = 0$, $a^{(2)}{}_6 = \Omega^{(0)}$, $a^{(2)}{}_7 = \Omega_1$, $a^{(2)}{}_8 = \Omega_2$, respectively. Correspondingly, we finally obtain the new symplectic matrix

$$(f_{ij}^{(2)}) = \begin{pmatrix} 0 & 0 & 0 & 0 & 0 & 0 & -1 & 0 \\ 0 & 0 & 0 & 0 & -1 & 0 & -1 & 0 \\ 0 & 0 & -2\partial_x & 0 & 0 & -c\partial_x & 0 & 0 \\ 0 & 0 & 0 & 2\partial_x & 0 & -c\partial_x & 0 & -\partial_x \\ 0 & 1 & 0 & 0 & 0 & -\partial_x & 0 & 0 \\ 0 & 0 & -c\partial_x & -c\partial_x & -\partial_x & 0 & 0 & 0 \\ 1 & 1 & 0 & 0 & 0 & 0 & 0 & 0 \\ 0 & 0 & 0 & -\partial_x & 0 & 0 & 0 & 0 \end{pmatrix} \delta(x-y). \quad (11)$$

Eq.(11) can be identified as a non-singular matrix, through very careful and very long calculation, its inverse is deduced as follows

$$(f_{ij}^{(2)})^{-1} =$$

$$\begin{pmatrix} -\frac{2}{c^2}\delta(x-y) & \frac{2}{c^2}\partial_x\delta(x-y) & \frac{1}{c}\delta(x-y) & 0 & -\delta(x-y) & -\frac{2}{c^2}\delta(x-y) & \delta(x-y) & \frac{2}{c}\delta(x-y) \\ \frac{2}{c^2}\partial_x\delta(x-y) & -\frac{2}{c^2}\partial_x\delta(x-y) & -\frac{1}{c}\delta(x-y) & 0 & \delta(x-y) & \frac{2}{c^2}\delta(x-y) & 0 & -\frac{2}{c}\delta(x-y) \\ -\frac{1}{c}\delta(x-y) & \frac{1}{c}\delta(x-y) & 0 & 0 & 0 & -\frac{1}{2c}\varepsilon(x-y) & 0 & \frac{1}{2}\varepsilon(x-y) \\ 0 & 0 & 0 & 0 & 0 & 0 & 0 & -\frac{1}{2}\varepsilon(x-y) \\ \delta(x-y) & -\delta(x-y) & 0 & 0 & 0 & 0 & 0 & 0 \\ \frac{2}{c^2}\delta(x-y) & -\frac{2}{c^2}\delta(x-y) & -\frac{1}{2c}\varepsilon(x-y) & 0 & 0 & \frac{1}{c^2}\varepsilon(x-y) & 0 & -\frac{1}{c}\varepsilon(x-y) \\ -\delta(x-y) & 0 & 0 & 0 & 0 & 0 & 0 & 0 \\ -\frac{2}{c}\delta(x-y) & \frac{2}{c}\delta(x-y) & \frac{1}{2}\varepsilon(x-y) & -\frac{1}{2}\varepsilon(x-y) & 0 & -\frac{1}{c}\varepsilon(x-y) & 0 & 0 \end{pmatrix}, \quad (12)$$

where $\varepsilon(x)$ is a general step-spring function, satisfying $\dfrac{d\varepsilon(x)}{dx} = 2\delta(x)$.

From which, we can identify the FJ generalized brackets[7] as

$$\{\xi_i^{(2)}(x), \xi_j^{(2)}(y)\}^* = f_{ij}^{(2)-1}(x, y) \quad . \quad (13)$$

And the quantization of gauge invariant self-dual fields is done by the usual replacement[7]

$$\{\xi_i^{(2)}(x),\xi_j^{(2)}(y)\}^* \to -\frac{i}{\hbar}[\hat{\xi}_i^{(2)}(x),\hat{\xi}_j^{(2)}(y)]. \qquad (14)$$

So far, we complete the FJ quantization of this system.

## 3 The comparison between FJ method and Dirac method in quantization of gauge invariant self-dual fields

Comparing the FJ generalized brackets (12) and (13) with the Dirac brackets[17], the relations of the two kinds of brackets are obtained as follows

$$\{A_1(x),\pi^1(y)\}^* = \{\xi_2(x),\xi_5(y)\}^* = f_{25}^{(2)-1}(x,y) = \delta(x-y) = \{A_1(x),\pi^1(y)\}_D \qquad ,\qquad (15)$$

$$\{A_1(x),A_1(y)\}^* = \{\xi_2(x),\xi_2(y)\}^* = f_{22}^{(2)-1}(x,y) = -\frac{2}{c^2}\partial_x\delta(x-y) = \{A_1(x),A_1(y)\}_D, \qquad (16)$$

$$\{A_0(x),A_1(y)\}^* = \{\xi_1(x),\xi_2(y)\}^* = f_{12}^{(2)-1}(x,y) = \frac{2}{c^2}\partial_x\delta(x-y) = \{A_0(x),A_1(y)\}_D \qquad ,\qquad (17)$$

$$\{A_0(x),A_0(y)\}^* = \{\xi_1(x),\xi_1(y)\}^* = f_{11}^{(2)-1}(x,y) = -\frac{2}{c^2}\partial_x\delta(x-y) = \{A_0(x),A_0(y)\}_D, \qquad (18)$$

$$\{A_0(x),\pi^1(y)\}^* = \{\xi_1(x),\xi_5(y)\}^* = f_{15}^{(2)-1}(x,y) = -\delta(x-y) = \{A_0(x),\pi^1(y)\}_D \qquad ,\qquad (19)$$

$$\{A_1(x),\phi(y)\}^* = \{\xi_2(x),\xi_3(y)\}^* = f_{23}^{(2)-1}(x,y) = -\frac{1}{c}\delta(x-y) = \{A_1(x),\phi(y)\}_D \qquad ,\qquad (20)$$

$$\{A_0(x),\phi(y)\}^* = \{\xi_1(x),\xi_3(y)\}^* = f_{13}^{(2)-1}(x,y) = \frac{1}{c}\delta(x-y) = \{A_0(x),\phi(y)\}_D \qquad ,\qquad (21)$$

$$\{A_0(x),\pi_\phi(y)\}^* = \{\xi_1(x),\partial_y\xi_3(y)\}^* = \partial_y f_{13}^{(2)-1}(x,y) = -\frac{1}{c}\partial_x\delta(x-y) = \{A_0(x),\pi_\phi(y)\}_D \quad ,$$

$$(22)$$

$$\{A_1(x),\pi_\phi(y)\}^* = \{\xi_2(x),\partial_y\xi_3(y)\}^* = \partial_y f_{23}^{(2)-1}(x,y) = \frac{1}{c}\partial_x\delta(x-y) = \{A_1(x),\pi_\phi(y)\}_D.$$

$$(23)$$

The other FJ generalized brackets and Dirac brackets all equal zero. We must emphasized that the FJ generalized brackets concerning $\xi_6$, $\xi_7$, $\xi_8$ are brought by multipliers, so there are no the correspondences to the Dirac brackets.

From the above discussion, it can be read that the FJ generalized brackets concerning real field variables and conjugate momenta coincide with the correspondent Dirac brackets. And this method in this letter can be applied to quantize many quantum systems in physics, e.g., 1+1 Dimensional Non-linear σ Model[18] etc, because the limit of the letter's space, the other more researches etc will be written in the other papers.

4 **Summary and Conclusion**

In this letter, we study gauge invariant self-dual fields interacting with gauge fields by using FJ method, obtain a new symplectic Lagrangian density, deduce the FJ generalized brackets of the gauge invariant self-dual fields interacting with gauge fields, further give Faddeev-Jackiw quantization of this system. By comparing FJ method with Dirac method for the model, we find that the FJ generalized brackets obtained by FJ method are identical with that obtained by Dirac method. Moreover, the resulting quantizations from these two methods are identical. So, we obtain the conclusion that the FJ method and Dirac method are equivalent in the quantization of gauge invariant self-dual fields interacting with gauge fields. By our practical research in this letter, we find that in contrast with Dirac method, the FJ method has the advantages of simplicity, obviates the need to distinguish primary and secondary constraints and first- and second-class constraints. Therefore, the FJ method is more economical and useful.